\documentclass[aps,reprint,twocolumn,pra]{revtex4-1}
\usepackage{graphicx,hyperref}
\usepackage{amsmath,amssymb}
\usepackage{xcolor,datetime}
\hypersetup{
    colorlinks=true,
    linkcolor=cyan,
    filecolor=cyan,      
    urlcolor=cyan,
    citecolor=cyan
}

\newcommand{\at}{\mathrm{a}}
\newcommand{\al}{\mathrm{a-l}}
\newcommand{\mem}{\mathrm{m}}
\newcommand{\ml}{\mathrm{m-l}}
\newcommand{\lf}{\mathrm{l}}
\newcommand{\Om}{\ensuremath{\Omega_\mem}}
\newcommand{\Gm}{\ensuremath{\Gamma_\mem}}
\newcommand{\Oa}{\ensuremath{\Omega_\at}}
\newcommand{\oL}{\ensuremath{{\omega_\mathrm{L}}}}
\newcommand{\aL}{\ensuremath{\alpha_\mathrm{L}}}
\newcommand{\oR}{\ensuremath{\omega_\mathrm{R}}}
\newcommand{\la}{\ensuremath{\lambda}}
\newcommand{\s}{\ensuremath{\sigma}}

\newcommand{\Hc}{\mathrm{H.c.}}
\renewcommand{\a}{\ensuremath{\alpha}}
\newcommand{\ga}{\ensuremath{\gamma}}
\newcommand{\crit}{\mathrm{c}}
\newcommand{\stwo}{\mathrm{s2}}
\newcommand{\sone}{\mathrm{s1}}
\newcommand{\aone}{\mathrm{a1}}
\newcommand{\NN}{\mathrm{N}}
\newcommand{\rme}{\mathrm{e}}
\newcommand{\bxi}{\boldsymbol{\xi}}
\newcommand{\ispi}{\int\frac{d\omega}{\sqrt{2\pi}}}
\newcommand{\abs}[1]{\ensuremath{\vert{#1}\vert}}
\newcommand{\ket}[1]{\ensuremath{\vert{#1}\rangle}}
\newcommand{\cmt}[2]{\ensuremath{[{#1},{#2}]}}
\newcommand{\mean}[1]{\ensuremath{\left\langle{#1}\right\rangle}}

\newcommand{\mtc}[1]{\ensuremath{\mathcal{#1}}}
\newcommand{\mbb}[1]{\ensuremath{\mathbb{#1}}}
\renewcommand{\Re}{\mathrm{Re}}
\let\oldPsi\Psi
\renewcommand{\Psi}{\hat{\oldPsi}}
\let\oldxi\xi
\renewcommand{\xi}{\hat{\oldxi}}

\begin{document}

\title[Tuning the Order of the NQPT in a Hybrid Atom-Optomechanical System]{Tuning the Order of the Nonequilibrium Quantum Phase Transition in a Hybrid Atom-Optomechanical System}
\author{N Mann$^1$, A Pelster$^2$, and M Thorwart$^1$}
\affiliation{$^1$I. Institut f\"ur Theoretische Physik, Universit\"at Hamburg, Jungiusstra\ss{}e 9, 20355 Hamburg, Germany \\
$^2$Physics Department and Research Center OPTIMAS, Technische Universit\"at Kaiserslautern, Erwin-Schr\"odinger Stra\ss{}e 46, 67663 Kaiserslautern, Germany}
\email{niklas.mann@physik.uni-hamburg.de}

\begin{abstract}
We show that a hybrid atom-optomechanical quantum many-body system with two internal
atom states undergoes both first- and second-order nonequilibrium quantum phase transitions. 
A nanomembrane is placed in a pumped optical cavity, whose outcoupled light forms a lattice for an ultracold Bose gas.
By changing the pump strength, the effective membrane-atom coupling can be tuned.
Above a critical intensity, a symmetry-broken phase emerges which is characterized by a sizeable occupation of the high-energy internal states and a displaced membrane.
The order of this nonequilibrium quantum phase transition can be changed by tuning the transition frequency.
For a symmetric coupling,  the transition is continuous below a certain transition frequency and discontinuous above.
For an asymmetric coupling, a first-order phase transition occurs.
\end{abstract}

\maketitle

\section{Introduction}
Using the concept of phase transitions, a great variety of different physical systems can be classified in terms of their emergent collective behaviour~\cite{Sachdev, Zinn-Justin,Kleinert}.
While phase transitions of both classical and quantum systems in equilibrium are by now quite well understood, the extension to nonequilibrium is a relatively new field.
In particular, it is of interest to understand which equilibrium properties survive at nonequilibrium, involving both external driving and dissipation.
On the other hand, novel properties may emerge in driven dissipative systems, where energy is not conserved, and the detailed balance condition and the fluctuation-dissipation theorem are no longer valid.
Yet, from an experimental point of view, it is not easy to realize and control systems with a nonequilibrium phase transition, in particular when quantum fluctuations dominate over thermal effects from the environment.
Currently discussed systems, which show nonequilibrium quantum phase transitions (NQPTs), are ultracold atoms in a lattice inside an optical cavity~\cite{Nagy2008,Maschler2008,Baumann2010,KlinderPRL,Bakhtiari2015} and
microcavity-polariton systems~\cite{Dagvadorj2015,Zamora2017,Comaron2018}.
Laser-driving offers the unique possibility to address and switch between
different phases of quantum many-body systems by tuning the pump strength.

Recently, also for hybrid atom-optomechanical quantum systems~\cite{Vogell2013,Joeckel2015,Zhong2017,Christoph2018,Vochezer2018}, a NQPT of second-order has been predicted~\cite{Mann2018}.
Such hybrids combine optomechanics with atom optics, as theoretically proposed~\cite{Vogell2013} and later experimentally realized~\cite{Joeckel2015,Zhong2017,Christoph2018,Vochezer2018}.
The vibrational motion of a nanomembrane in an optical cavity is coupled to the spatial motion of a distant cloud of cold $^{87}$Rb atoms that reside in the optical lattice of the outcoupled light field.
By combining different cooling mechanisms such as optical feedback cooling~\cite{Christoph2018} and sympathetic cooling by utilizing the atom gas as a coolant~\cite{Vogell2013,Joeckel2015,Zhong2017,Christoph2018}, the nanooscillator can be cooled close to its quantum mechanical ground state.
Quantum many-body effects lead to collective atomic motion with an instability~\cite{Vochezer2018} and a second-order NQPT~\cite{Mann2018} to a state with a spatially shifted optical lattice.
Besides, indirect quantum measurement, atom-membrane entanglement and coherent state transfer are in the focus of interest~\cite{Hammerer2009a,Hammerer2009b,Wallquist2009,Paternostro2010,Genes2011}.

A significant drawback in the motional coupling scheme~\cite{Vogell2013} is the strong frequency mismatch between the nanooscillator and the atomic motion in the optical trap which hinders resonant coupling.
A decisive advance is the use of internal atomic states instead of their spatial degree of freedom, such that this internal state coupling scheme~\cite{Vogell2015} enables resonant coupling.
Here, the motion of the mechanical membrane is indirectly coupled to transitions between internal states of the atoms via translating the phase shift of the light, caused by the membrane displacement, into a polarization rotation using a polarizing beam splitter (PBS).
By this scheme, membrane cooling~\cite{Vogell2015,Lau2018}, or a displacement-squeezed membrane~\cite{Ockeloen2013} can be realized.
The atoms can implement an effective harmonic oscillator with negative mass~\cite{Polzik2015}, that, in turn, can be utilized for quantum back-action evading measurements~\cite{Moller2017}, enabling a high displacement sensitivity.
Moreover, the collective nature of the hybrid system mediates long-range interactions in the atom gas, similar to those in a spinor dipolar Bose-Einstein
condensate~\cite{Yi2006,Kawaguchi2006}.

In this work, we show that the internal state coupling scheme also allows for a NQPT, whose order can be readily tuned by changing the atomic transition frequency.
Thus, both a first- and a second-order NQPT can be realized in the same physical set-up by only changing a directly accessible control parameter.
We show this for the “membrane-in-the-middle-setup”~\cite{Vogell2015}, where the adiabatic elimination of the light field yields an effective coupling between the membrane and the transition between two states in the atom gas, see Fig.~\ref{fig1}.
In a mean-field description, the atomic part is reduced to a single-site problem with a Gaussian ansatz for the condensate profile. 
Tuning the atom-membrane coupling by changing the laser intensity, the system undergoes a NQPT.
We provide simple analytical expressions for the resulting critical point.
Moreover, by tuning the atomic transition frequency, even the order of the phase transition can be changed from second- to first-order and \textit{vice versa}.
In case of a discontinuous phase transition, the system exhibits a characteristic hysteresis which can be detected by measuring the occupation of the internal states of  atom gas.
Throughout this work, we assume natural units and consequently set $\hbar=c=1$.

\section{Model and Adiabatic Elimination of the Light Field}
\begin{figure*}
\centering\includegraphics[width=0.75\textwidth]{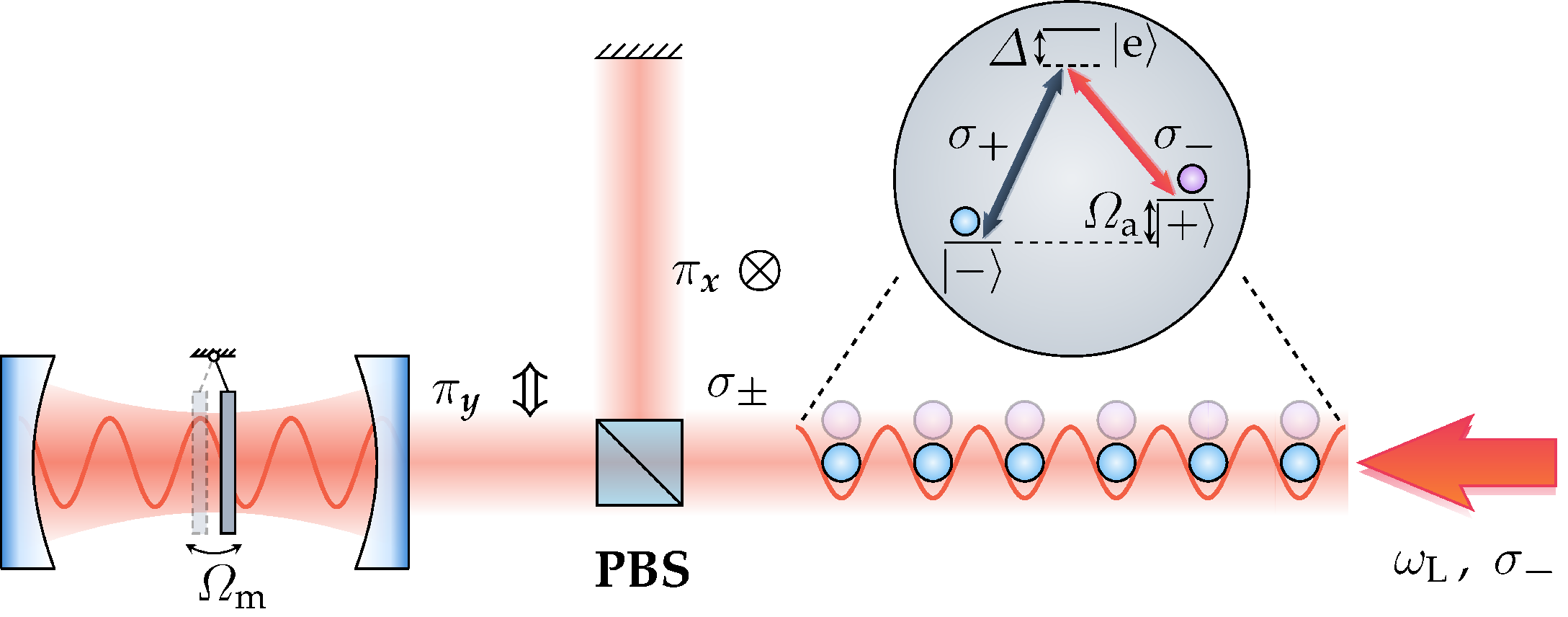}
\caption{A nanomechanical membrane in an optical cavity is coupled to the internal states of a distant atomic ensemble that is trapped in an optical lattice. The internal states of the atoms are arranged in a $\Lambda$-type scheme according to the inset.}
\label{fig1}
\end{figure*}

We consider an ensemble of $N$ ultracold $^{87}$Rb atoms placed in an external optical lattice.
The atoms exhibit three relevant internal states $\tau\in\{+,-,\rme\}$ that are arranged in a $\Lambda$-type level scheme.
The two low-energy states are energetically separated by the atomic transition frequency \Oa, which can be tuned by an external magnetic field.
The transition between the states \ket{+} and \ket{\rme} is driven by an applied $\s_-$ circularly polarized laser with frequency \oL.
The passing beam is directed to a PBS, which divides the circularly polarized light into linearly polarized $\pi_x$ and $\pi_y$ light beams on two perpendicular arms, which are of equal length measured for an undisplaced membrane, see figure~\ref{fig1}.
The vertical path involves a fixed mirror which reflects light with conserved polarization $\pi_x$. In the horizontal path, a nanomembrane with resonance frequency \Om\ is placed inside a low-finesse cavity, which reflects $\pi_y\rightarrow\pi_y$ light when undisplaced. 
The light of both arms is directed back onto the atoms mediating the effective atom-membrane coupling.

In a quasi-static picture, a finite displacement of the membrane induces a finite phase shift on the propagating horizontal $\pi_y$ beam leading to a rotated polarization after the light has passed the PBS backwards.
The emergent $\s_+$ photon now impinges on an atom and may induce a two-photon transition between the states \ket{-}$\leftrightarrow$\ket{+}, when the resonance condition $\Om\simeq\Oa$ is met.
The back-action of the atoms on the membrane is induced by a transition of the atoms between the states \ket{-} and \ket{+}.
The emitted $\s_+$  photons pass the PBS with 50\% probability horizontally and hit the membrane. 
This changes the radiation pressure on the membrane.

The prototypical system Hamiltonian can be written as a sum of five consecutive terms
\begin{equation}
\hat{H}_\mathrm{tot} = \hat{H}_\at + \hat{H}_\mem + \hat{H}_\lf + \hat{H}_{\al} + \hat{H}_{\ml}\,. \label{eq:Hlin}
\end{equation}
Each of the first three terms describes one of the three compounds: the atomic condensate, the nanomembrane, and the light field, respectively.
The atom-light field interactions and the optomechanical coupling are included in $\hat{H}_\al$ and $\hat{H}_\ml$, respectively.
We consider one well separated vibrational mode of the membrane which is modeled as a single harmonic oscillator
\begin{equation} 
\hat{H}_\mem = \Om \hat{a}^\dag \hat{a}\,,
\end{equation}
with the mechanical frequency \Om\ and bosonic annihilation (creation) operator $\hat{a}$ ($\hat{a}^\dag$) which follow the usual bosonic commutation relation $\cmt{\hat{a}}{\hat{a}^\dag}=1$.
The atomic gas is modeled by the standard many-body Hamiltonian
\begin{align}
\hat{H}_\at = \sum_{\tau=\pm}\int d z\, \Psi_\tau^\dag(z)\Biggl[&- \frac{\partial_z^2}{2m} +\frac{1}{2}\sum_{\tau'=\pm}g_{\tau\tau'} \Psi_{\tau'}^\dag(z)\Psi^\dag_{\tau'}(z) \nonumber\\
&+V_\tau(z)\Biggr]\Psi_\tau(z)\,,
\end{align}
where the potential $V_\tau(z)$ includes the energy of the corresponding internal state as well as an optical potential, and $m$ is the mass of an atom.
Moreover, we assume a contact interaction with the one-dimensional interaction strength $g_{\tau\tau'}$, which, in general, can be different depending on the internal states in which the atoms reside.
We assume a large detuning $\Delta$ between the frequency of the pump laser and the addressed transition, such that we can eliminate the excited state \ket{\rme}.
Hence, we only consider the two internal states \ket{-} and \ket{+} in our description.
The light modes have two possible optical polarizations $\s_-$, $\s_+$ which are represented by the bosonic operators $\hat{b}_{\omega-}$, $\hat{b}_{\omega+}$, respectively, and are included over a bandwidth $2\theta$ around the laser frequency \oL.
They are described by
\begin{equation}
\hat{H}_\lf = \int_{\oL-\theta}^{\oL+\theta} d\omega\,\omega\left(  \hat{b}^\dag_{\omega-}\hat{b}_{\omega-} + \hat{b}^\dag_{\omega+}\hat{b}_{\omega+}\right).
\end{equation}

\subsection{Linearized Coupling of the Membrane and the Light Field}
The external pumping laser has a $\s_-$ polarization, such that the coherent drive is included by the linear replacement at the laser frequency \oL
\begin{equation}
\hat{b}_{\omega-}\rightarrow \hat{b}_{\omega-} + \delta(\omega-\oL)e^{-i\oL t}\aL\,.
\end{equation}
In the following, we assume $\abs{\aL}\gg1$ such that the interaction between the light field and the membrane (atoms) can be linearized in the operators $\hat{b}_{\omega-}$ and $\hat{b}_{\omega+}$.
In a reference frame that rotates with the laser frequency \oL, the linearized membrane-light field interaction takes the form~\cite{Vogell2015}
\begin{equation}
\hat{H}_\ml = \la_\mem \left( \hat{a}+\hat{a}^\dag \right) \ispi \left( \hat{b}_{\omega-} + \hat{b}^\dag_{\omega-} + \hat{b}_{\omega+} + \hat{b}^\dag_{\omega+}\right)
\end{equation}
with the coupling strength $\la_\mem$.
In the "membrane-in-the-middle" setup, the membrane-light field coupling $\la_\mem = 2{\aL\abs{r_\mem}\oL\ell_\mem} {\mathcal{F}}/{\pi^{3/2}}$ scales with the cavity finesse \mtc{F} and the light field amplitude $\aL$.
Moreover, \abs{r_\mem} is the membrane reflectivity and $\ell_\mem=(M\Om)^{-1/2}$ denotes the amplitude of the zero-point motion of the membrane, where $M$ stands for the mass of the membrane~\cite{Vogell2013}.
Here, we have neglected the quadratic term in \aL, which leads to a constant linear force on the membrane and, thus, only alters its equilibrium position.
This can be accounted for by a simple redefinition of the zero-point position.

The dipolar interaction of the atoms with the light field induces an AC-Stark shift of the electronic levels of the atoms.
After the elimination of the auxiliary excited state~\ket{\rme} and the linearization in the light field operators, the atom-light field coupling is given by
\begin{align}
\hat{H}_\al =& \ispi \int\! d z \sin(\oL z) \sin(\omega z)[\la_\at\hat{b}_{\omega-}\Psi_+^\dag(z)\Psi_+(z)\nonumber\\&+\la_\pm\hat{b}_{\omega+}\Psi_+^\dag(z)\Psi_-(z)+\Hc]\,,\label{eq:Hal}
\end{align}
which includes two essentially different processes.
On the one hand, the first line couples atoms in the internal state~\ket+ to the photon field quadrature in a fashion similar to the motional coupling scheme~\cite{Vogell2013}.
This interaction emerges due to the driving of the atomic transition $\ket+\leftrightarrow\ket\rme$ and scales according to $\la_\at=\sqrt{2\pi} \aL\oL\mu_+^2 \mathcal{E}_\oL/\Delta$, where $\mu_+$ is the transition dipole element of the corresponding transition and $\mathcal{E}_\oL=\sqrt{\oL/\pi\mathcal{A}}$ is a normalization constant of the light mode operators with \mtc{A} being the beam cross-sectional area.
On the other hand, the second line of Eq.~\eqref{eq:Hal} includes transitions between the atomic internal states under the creation (or annihilation) of a $\s_+$ polarized photon.
Similarly, this interaction constant is given by $\la_\pm=\sqrt{2\pi} \aL\oL\mu_+\mu_- \mathcal{E}_\oL/\Delta$, where $\mu_-$ is now the atomic transition dipole moment between the two states $\ket-\leftrightarrow\ket\rme$.

In order to simplify the linearized atom-light field coupling, the external potential $V_\tau(z)$ can be chosen such that the atoms are positioned around the lattice sites $z_j$, defined by the relation $\sin(2\oL z_j)=1$.
An additional potential for the atoms in the state \ket+ has to be provided in order to cancel the lattice potential generated by the coherent drive.
This leads to a constant term that redefines the atomic transition frequency.
Overall, we choose the potential according to $V_\tau(z)=\tau\Oa/2 + V\cos^2 \oL z$, where $V$ characterizes the lattice depth.

\subsection{Adiabatic Elimination of the Light Field and Effective Equations of Motion}
In the following, we consider an optical cavity in the bad cavity regime in order to adiabatically eliminate the light field modes.
We assume that the cavity linewidth  is large compared to both the atomic and the membrane frequency such that both sideband photons are well accommodated in the response profile of the cavity.
To do so, we start with the linearized Hamiltonian of Eq.~\eqref{eq:Hlin} in the interaction picture with respect to the light field Hamiltonian $\hat{H}_\lf$.
Hence, the light mode operators transform according to $\hat{b}_{\omega\mu}(t)_{\rm I} = \hat{b}_{\omega\mu}\exp[-i(\omega-\oL)t]$, where the index ${\rm I}$ labels the interaction picture.

The formal solution of the Schr\"odinger equation for any arbitrary state \ket\psi\ in the interaction picture reads
\begin{equation}
\ket{\psi(t)_{\rm I}} = \mtc{T} \exp\left\{-i\int_0^t d s\, \hat{H}'(s)_{\rm I}\right\}\ket{\psi(0)}\,,
\end{equation}
with the time-ordering operator \mtc{T} and $\hat{H}' = \hat{H}_{\rm tot}-\hat{H}_\lf$.
Next, we expand the equation on the right-hand side for small time steps $\delta t$.
Up to second order, the relevant terms read
\begin{align}\begin{split}
\ket{\psi(\delta t)_{\rm I}} \simeq& \Biggl\{1 - i\int_0^{\delta t} dt\, \hat{H}'(t)_{\rm I} \\&- \int_0^{\delta t} dt \int_0^t ds\, \hat{H}'(t)_{\rm I}\hat{H}'(s)_{\rm I}\}\Biggr\}\ket{\psi(0)}\,.\label{eq:QSSE}
\end{split}\end{align}
Moreover, we assume that the initial state is a product state $\ket{\psi(0)} = \ket\psi_{\at+\mem}\otimes\ket{\mathrm{vac}}_\lf$, where $\ket{\mathrm{vac}}_\lf$ denotes the vacuum state of the light field and $\ket{\psi}_{\at+\mem}$ stands for an arbitrary state in the atom-membrane subspace.
Under these assumptions, the photon mode operators fulfill $\hat{b}_{\omega\mu}\ket{\psi(0)}=0$ and we may define the noise-increment operators
\begin{align}
\delta \hat{B}(t) &= \int_{t}^{t+\delta t} ds\, \ispi \Bigl[\hat{b}_{\omega+}(s)_{\rm I} + \hat{b}_{\omega-}(s)_{\rm I}\Bigr]\,,\\
\delta \hat{C}_\mu(t,z) &= \int_t^{t+\delta t} ds\, \ispi \sin(\oL z) \sin(\omega z) \hat{b}_{\omega\mu}(s)_{\rm I} \,.
\end{align}

Finally, we take the limit $\delta t\rightarrow0$ and assume that the noise-increment operators after a time step $\delta t$ do not depend on their form at an earlier time.
This assumption is equivalent to the Markov approximation.
With this, we can derive a quantum stochastic Schr\"odinger equation (QSSE) in the Ito form~\cite{ Vogell2015,bookGardiner} with $d\ket{\psi(t)} = \ket{\psi(t+dt)} - \ket{\psi(t)}$.
The differential noise operators, e.g., $\lim_{\delta t\rightarrow0}\delta \hat{B}(t)= d\hat{B}(t)$, follow the Ito rules
\begin{align}
d\hat{B}(t) d\hat{B}^\dag(t) =& 2dt\,,\\
d\hat{B}(t) d\hat{C}_\mu^\dag(t,z) =& \sin(\oL z) dt\,,\\
 d\hat{C}_\mu(t,z)d\hat{B}^\dag(t) =& \sin(\oL z) dt\,,\\
d\hat{C}_\mu(t,z) d\hat{C}_\mu^\dag(t,z') =& \sin(\oL z) \sin(\oL z') dt\,.
\end{align}
Under the consideration of these relations, we can derive the quantum Langevin equation from the QSSE~\eqref{eq:QSSE} which describes the dynamics of the membrane mode operator $\hat{a}$ and the atomic field operators $\Psi_\tau$.
The hybrid atom-membrane system is then effectively described by the equations of motion
\begin{widetext}\begin{align}
i\partial_t \hat{a} = & \left(\Om - i\Gm\right)\hat{a} - \la\chi \int dz\, \cos(2z)\Psi^\dag_+(z)\Psi_+(z) -\frac{\la}{2}\int dz\, \cos(2z) \left[\Psi_+^\dag(z)\Psi_-(z) + \Hc\right] + \xi_\mem \label{eq:Langevin1}\,,\\
i\partial_t \Psi_-(z) =& \left[-\frac{\Oa}{2} - \oR \partial_z^2 - \frac{V}{2}\cos(2z) + \sum_{\tau=\pm} g_{\tau-}\Psi_\tau^\dag(z)\Psi_\tau(z)\right] \Psi_-(z)- \frac{\la}{2}\left(\hat{a} + \hat{a}^\dag\right) \cos(2z) \Psi_+(z)\,,\\
i\partial_t \Psi_+(z) =& \left[\frac{\Oa}{2} - \oR \partial_z^2 - \biggl(\frac{V}{2}+\la\chi\bigl\{\hat{a}+\hat{a}^\dag\bigr\}\biggr)\cos(2z) + \sum_{\tau=\pm} g_{\tau+}\Psi_\tau^\dag(z)\Psi_\tau(z)\right] \Psi_+(z)- \frac{\la}{2}\left(\hat{a} + \hat{a}^\dag\right) \cos(2z) \Psi_-(z)\,,\label{eq:Langevin2}
\end{align}\end{widetext}
where we have scaled and shifted the atom position variable $\oL z \rightarrow z + \pi/2$, such that the lattice minima for $V>0$ are located at the position $z_j = j\pi$ with $j\in\mbb{Z}$.
Here, we have defined the atom-membrane coupling constant $\la=\la_\mem \la_\pm /2$, which corresponds to the process that induces transitions between the internal states under the creation (annihilation) of a phonon, and $\oR=\oL^2/2m$ is the atomic recoil frequency.
Moreover, the coupling of the membrane to the number of atoms in the internal state \ket+\ is given by $\la'=\la_\mem \la_\at/2$ .
The latter, in fact, is not independent of the internal state coupling constant \la\ as $\la'/\la=\chi = \mu_+/\mu_-$, such that we can choose the parametrization $\la'=\la\chi$.
In addition, we have neglected terms introduced by the light field that lead to long-range interactions in the atom gas.
This assumption is justified if the laser frequency $\oL$ is far detuned from the transition frequency between the states \ket+ and \ket\rme.
Finally, fluctuations introduced by the light field have been neglected.
A phenomenological damping of the membrane mode has been introduced with rate \Gm\ together with the corresponding bosonic noise operator $\xi_\mem$ that is characterized by the autocorrelation functions 
\begin{align}\begin{split}
\mean{\xi_\mem(t) \xi_\mem^\dag(0)}=&2\Gm(N_\mem+1)\delta(t)\,,\\
\mean{\xi_\mem^\dag(t) \xi_\mem(0)}=&2\Gm N_\mem\delta(t)\,.\end{split}\label{eq:autocorr}
\end{align}
Here, $N_\mem$ is the environment occupation number which determines the steady-state occupation of the phonon mode in the isolated limit $\la=0$.

\section{Tuning the Order of the Quantum Phase Transition}
Assuming that the atoms are prepared at low temperature and for a weak atom-membrane coupling such that a large fraction of atoms occupies the ground state and a condensate is formed, the combined system dynamics is subject to the set of coupled mean-field equations of motion
\begin{widetext}\begin{align}
 i\partial_t\a =& \left(\Om - i\Gm\right)\a - \sqrt{N}\la \int dz\, \cos(2z)\left[\chi\abs{\psi_+}^2 + \mathrm{Re}(\psi_+^*\psi_-)\right]\,,\label{eq:eom-mem}\\
i\partial_t \psi_- =& \left[-\frac{\Oa}{2}-\partial_z^2\oR -\frac{V}{2}\cos(2z) + N \sum_{\tau=\pm} g_{\tau-}\abs{\psi_\tau}^2\right]\psi_-- {\sqrt{N}\la}\Re(\a)\cos(2z)\psi_+\,,\label{eq:eom-psim}\\
i\partial_t \psi_+ =& \left[\frac{\Oa}{2}-\partial_z^2\oR -\frac{V}{2}\cos(2z) + N \sum_{\tau=\pm} g_{\tau+}\abs{\psi_\tau}^2\right]\psi_+ - {\sqrt{N}\la}\Re(\a)\cos(2z)\psi_- - {2\sqrt{N}\la\chi}\Re(\a)\cos(2z)\psi_+ \label{eq:eom-psip}\,.
\end{align}\end{widetext}
Here, the first equation describes the motion of the membrane, and the second and third equation describe the dynamics of the atomic condensate in the internal state \ket- and \ket+, respectively.
The complex amplitude $\a=\mean{\hat{a}}/\sqrt{N}$ is the scaled mean value of the ladder operator $\hat{a}$ and $\psi_\pm=\mean{\Psi_\pm}/\sqrt{N}$ is the condensate wave function of the corresponding internal atom state, where $N$ denotes the total number of atoms.

\subsection{Single Mode Approximation and Cumulant Expansion}
For a sufficiently deep lattice $V\gg\oR$, the condensate profile is well described by a sum of Gaussians residing in the individual lattice wells. 
When the wave function overlap between neighboring sites is small, the problem reduces to an effective single-site problem.
It is then reasonable to make the ansatz
\begin{equation}
\psi_\tau(t,z) = \ga_\tau \left(\frac{1}{\pi\s_\tau^2}\right)^{1/4} \exp\left(-\frac{z^2}{2\s_\tau^2}+i\eta_\tau z^2\right)\,,\label{eq:Gauss}
\end{equation}
with a constant number of atoms, i.e., the occupation amplitudes $\ga_\pm(t)$ fulfill the condition $\abs{\ga_-(t)}^2+\abs{\ga_+(t)}^2=1$, the individual condensate widths $\s_\tau(t)$ and the corresponding phases $\eta_\tau(t)$ which are used to induce the dynamics for $\s_\tau(t)$.
In order to reduce the number of parameters, we restrict ourselves to the special case $g_{\tau\tau'}=g$.
Finally, let us note that a mixing of the condensate profiles for different internal states is the energetically preferred state.
Already from equation \eqref{eq:eom-mem} it can be concluded that a maximally mixed condensate maximizes the effective coupling between the atoms and the membrane. 
This will eventually lead to a minimization of an effective nonequilibrium potential, which we will derive in the following.

\begin{figure*}
\centering\includegraphics[width=0.85\textwidth]{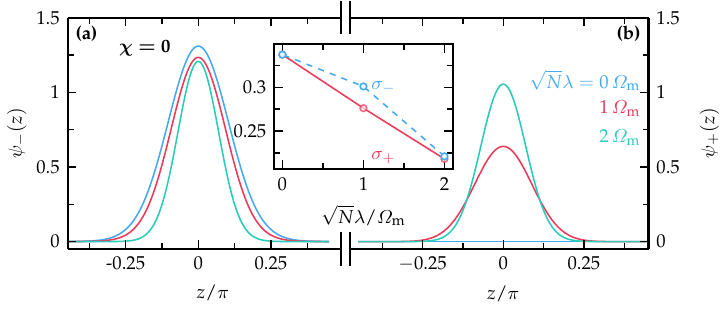}\\
\centering\includegraphics[width=0.85\textwidth]{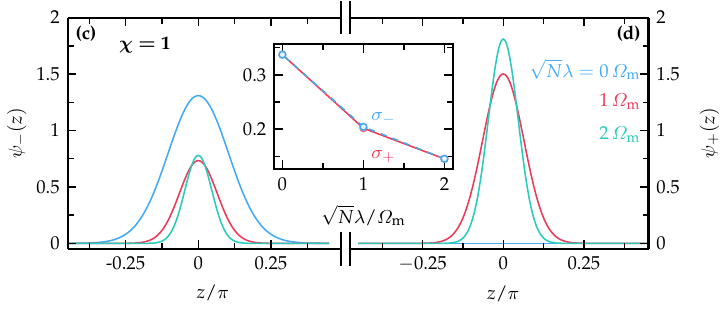}
\caption{ The steady-state condensate profile for the internal state \ket-\ in (a), (c) and \ket+\ in (b),(d) is shown for different values of the atom-membrane coupling, as indicated in panel (b), (d). 
A symmetric coupling with $\chi = 0$ is considered in (a) and (b), while the asymmetric case for $\chi=1$ is shown in panels (c) and (d).
The inset shows the condensate profile width of the corresponding internal state obtained via a Gaussian fit to the condensate profile. 
Other parameters used are $V = 100\,\oR$, $Ng = \oR$, $\Oa = 50\,\oR$, $\Om = 100\,\oR$, and $\Gm = 10\,\oR$.
}
\label{fig2}
\end{figure*}

In order to justify the ansatz, we numerically determine the steady
state of the extended Gross--Pitaevskii equation (GPE) \eqref{eq:eom-mem}--\eqref{eq:eom-psip} by using an imaginary time evolution with the Crank--Nicolson scheme.
Due to the periodicity of the potential, we use periodic boundary conditions and evaluate the steady state within the interval from $-\pi/2$ to $\pi/2$.
The condensate profile around a single potential well is shown for the symmetric coupling case $\chi=0$ in Fig.~\ref{fig2}(a) and (b) and the asymmetric case $\chi=1$ in Fig.~\ref{fig2}(c) and (d).
Here, different coupling constants \la\ have been chosen according to the color coding in Fig.~\ref{fig2}(b).
The panels (a), (c) and (b), (d) show the condensate profile of the atoms in the internal states \ket{-} and \ket{+}, respectively, as a function of the position coordinate $z$.
The well minimum is located at $z=0$.
Moreover, the insets compare the individual widths $\s_-$ and $\s_+$ obtained from a Gaussian fit to the condensate profile according to equation \eqref{eq:Gauss}.
The deviations between the individual widths are negligible in most cases and slightly increase only in the vicinity of a certain critical point.
Consequently, we can approximate the condensate profiles by a unified condensate width $\s\equiv \s_-=\s_+$ and an equal phase $\eta\equiv\eta_-=\eta_+$.

Next, we perform a cumulant expansion~\cite{Mann2018,Busch1997} of the equations of motion in order to determine the dynamics of the respective variational parameters.
Thus, we calculate the (i) zeroth and (ii) second cumulants by multiplying equations \eqref{eq:eom-psim} and \eqref{eq:eom-psip} (i)~with $\psi_0^*(z)=\exp(-z^2/2\s^2 - i\eta z^2)/(\pi\s^2)^{1/4}$ as well as (ii) with $(z^2-\s^2/2)\psi^*_0(z)$ and integrate then over $z$.
This leads to five independent equations of motion of which one is given by $\dot{\s}=4\oR\eta\s$.
By defining the effective potential
\begin{align}
 E &= \Om\abs{\a}^2 + \frac{\Oa}{2}\left(\abs{\ga_+}^2-\abs{\ga_-}^2\right) + \frac{\oR}{2\s^2} - \frac{V}{2}e^{-\s^2}  \nonumber\\
  &+ \frac{Ng}{\sqrt{8\pi}\s}- \sqrt{N}\la\left(\a+\a^*\right)\left(\chi\abs{\ga_+}^2 + \mathrm{Re}[\ga_+^*\ga_-]\right)e^{-\s^2}\,,\label{eqn:Epot}
\end{align}
the remaining equations of motion are given in a compact form as 
\begin{eqnarray}
\dot{\a} = -i\partial_{\a^*} E - \Gm \a\,,\label{eq:eom-cum1}\\
\dot{\ga}_\tau = -i\partial_{\ga_\tau^*} E\,,\\
(4\oR)^{-1} \ddot{\s} = -\partial_\s E\,.\label{eq:eom-cum2}
\end{eqnarray}
We note that the last term in equation \eqref{eqn:Epot} reflects the argument that a maximally mixed atomic condensate minimizes the effective potential energy.

\subsection{Nonequilibrium Potential and Steady-State Configuration}
In the presence of damping, the system will eventually relax to a steady non-thermal state.
Thus, each of the parameters can be split into its steady-state value and deviations from the steady state.
In this section, we are mainly interested in the steady-state properties of the combined hybrid system.
That is, we make the ansatz $\a(t)=\a_0$, $\ga_-(t) = \sqrt{1-\ga_0^2}$, $\ga_+(t)=\ga_0$ and $\s(t)=\s_0$ with a real-valued polarization $\ga_0$.
By inserting this ansatz in equations~\eqref{eq:eom-cum1}--\eqref{eq:eom-cum2}, the relation for the membrane amplitude
\begin{equation}
\a_0(\ga,\s) = \frac{\sqrt{N}\la }{\Om - i\Gm}\left( \chi\ga^2 + \ga\sqrt{1-\ga^2}\right) e^{-\s^2}\label{eqn:a0}
\end{equation}
is found.
With this, the effective potential of Eq.~\eqref{eqn:Epot} can be expressed in terms of the condensate variational parameters as 
\begin{align}
 E[\ga,\s] =& \frac{\Oa}{2}\left(2\ga^2-1\right) - \frac{N\la^2}{\Om'e^{2\s^2}}\left(\chi\ga^2 + \ga\sqrt{1-\ga^2}\right)^2 \nonumber\\
 &+ \frac{\oR}{2\s^2} - \frac{V}{2e^{\s^2}} + \frac{Ng}{\sqrt{8\pi}\s}\,,\label{eqn:nEpot}
\end{align}
which includes the nonequilibrium condition in the form of the primed mechanical frequency $\Om' = \Om + \Gm^2/\Om$.

For a qualitative understanding of the role of both the atom-membrane coupling \la\ and the asymmetry $\chi$, we study the potential surface of equation~\eqref{eqn:nEpot}.
Now, the global minimum of the effective nonequilibrium potential $E[\ga_0,\s_0]=E_0$ defines the steady-state configuration $\ga_0$, $\s_0$ and accordingly via Eq.~\eqref{eqn:a0} also $\a_0$.
The defining equations for the steady-state values of the condensate variational parameters are given by
\begin{align}\begin{split}
 \frac{4N\la^2}{\Om'}\left( \chi\ga_0^2 + \ga_0 \sqrt{1-\ga_0^2}\right)^2e^{-2\s_0^2} = & \frac{\oR}{\s_0^4} - V e^{-\s_0^2} \\ &+ \frac{Ng}{\sqrt{8\pi}\s_0^3}\,,\end{split}\label{eq:s0}\end{align}
\begin{align}\begin{split}
 \ga_0\Biggl(\frac{\la_\Omega}{\la}e^{\s_0^2}\Biggr)^2=&\ga_0\Bigl(\chi\ga_0 + \sqrt{1-\ga_0^2}\Bigr)\\
 &\times\Bigl(2\chi\ga_0\sqrt{1-\ga_0^2} + 1-2\ga_0^2\Bigr) \,,\end{split}\label{eq:ga0}
\end{align}
where we have introduced the coupling rate $\sqrt{N}\la_\Omega = \sqrt{\Oa\Om'}$.
While Eq.~\eqref{eq:s0} has only one possible solution $\s_0$ for a given occupation amplitude $\ga_0$, Eq.~\eqref{eq:ga0} allows, in general, multiple steady-state configurations $\ga_0$ that minimize the effective potential \eqref{eqn:nEpot}.
Rather than minimizing the potential with respect to all three parameters, we minimize it with respect to $\a_0$, $\s_0$ for a given occupation amplitude $\ga$, which is taken as an order parameter, and study the resulting potential energy surface $E(\ga) = E[\ga, \s_0(\ga)]$.
The global symmetry properties of the hybrid system are then determined by $\ga$ via the
influence of the occupations of the condensate species.

\begin{figure*}
\centering\includegraphics[width=\textwidth]{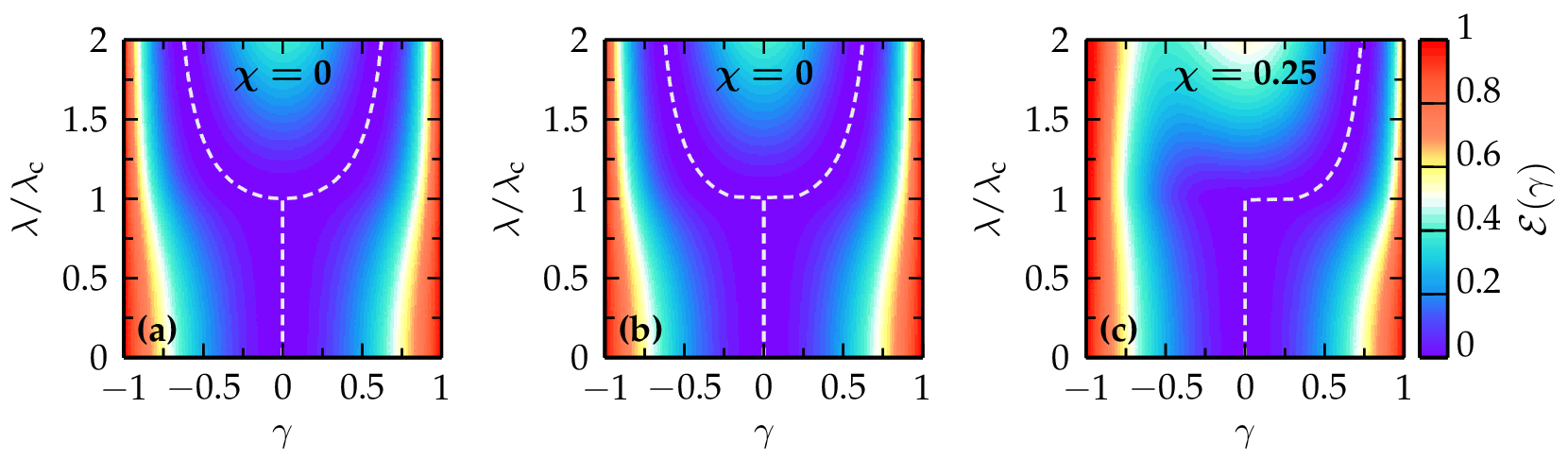}
\caption{The normalized potential surface $\mtc{E}(\ga)$ is shown as a function of the atomic occupation parameter $\ga$ and the atom-membrane coupling strength $\la$.
The dashed white curves show the steady-state configuration $\ga_0$.
In panel (a) and (b), a symmetric coupling ($\chi=0$) is considered with $\Oa= 50\,\oR$ and $\Oa= 5000\,\oR$, respectively.
In (a), the system exhibits a second-order phase transition, whereas the system undergoes a first-order phase transition in (b).
In panel (c), the asymmetric coupling regime is considered with $\chi = 0.25$ and $\Oa= 50\,\oR$.
Here, the system always exhibits an asymmetric first-order phase transition, characterized by a single favored atomic polarization. The other parameters throughout all panels are $V = 100\,\oR$, $Ng = \oR$, $\Om = 2\,\Oa$ and $\Gm = 0.1\,\Om$.
}
\label{fig3}
\end{figure*}
Figure~\ref{fig3} shows the resulting normalized energy surface $\mtc{E}(\ga) = [E(\ga)-E_0]/\max_{\abs{\bar{\ga}}\leq1}[E(\bar{\ga})-E_0]$.
The dashed curves indicate the occupation amplitude $\ga_0$ that globally minimizes the energy potential.
Below a certain coupling value $\la\leq\la_\crit$, the potential is minimized for $\ga_0=0$ in which all atoms occupy the state \ket{-}.
At $\la=\la_\crit$, the system undergoes a NQPT that is characterized by a non-vanishing occupation amplitude $\ga_0\neq0$ with different characteristics.
The case of symmetric coupling $\chi=0$ is shown in Fig.~\ref{fig3}(a) and Fig.~\ref{fig3}(b) for the transition frequency $\Oa=50\,\oR$ and $\Oa=5000\,\oR$, respectively.
In Fig.~\ref{fig3}(a), the NQPT is of second-order with a critical behavior $\ga_0\sim(\la-\la_\crit)^{1/2}$.
Interestingly, by tuning the atomic transition frequency $\Oa$, the NQPT becomes a symmetric first-order phase transition where the order parameter shows a jump at the critical point, see Fig.~\ref{fig3}(b).
Here, the bistable phase corresponds to the two states $(\ket{-}\pm\ket{+})/\sqrt{2}$.
For a non-vanishing asymmetry $\chi>0$, even an asymmetric first-order phase transition occurs at a critical coupling, where the left branch $(\ket{-}-\ket{+})/\sqrt{2}$ is energetically raised, such that the right branch $(\ket{-}+\ket{+})/\sqrt{2}$ represents the minimum, see Fig.~\ref{fig3}(c).

In the case of a second-order NQPT, we label the critical coupling by $\la_\stwo$.
An implicit definition of $\la_\stwo$ is found by inserting the steady-state solution of the condensate width $\s_0^2 = \log(\la_\stwo/\la_\Omega)$ of equation~\eqref{eq:ga0} for $\ga_0=0$ in the equation~\eqref{eq:s0}.
Hence, we find the implicit equation for the critical coupling rate
\begin{equation}
\oR + \frac{Ng}{\sqrt{8\pi}}\sqrt{\log\frac{\la_\stwo}{\la_\Omega}} = V \left(\frac{\la_\Omega}{\la_\stwo}\right) \left(\log\frac{\la_\stwo}{\la_\Omega}\right)^2\,.
\end{equation}
Yet, in the event of a first-order NQPT, such an implicit definition of the corresponding critical coupling rate can not be found on the basis of a set of steady-state equations.
However, a procedure to find the critical points can be defined by performing a Landau expansion of the effective nonequilibrium potential $E(\ga)$.
Moreover, Landau theory allows us to classify the order of the phase transition by evaluating the expansion coefficients.

\subsection{Landau Expansion of the Nonequilibrium Potential}
In order to verify whether Fig.~\ref{fig3}(b) indeed shows a first-order phase transition, we expand the nonequilibrium potential $E(\ga)$ in the order parameter around $\ga_0=0$.
Due to the asymmetry in the coupling, the Taylor expansion takes in general the form $E(\ga) = a_0 + \sum_{n\geq2} a_n \ga^n$, allowing also odd orders in $n$.
In order to fix the condensate width to its value $\s_0(\ga)$, we define the auxiliary function
\begin{align}
F[\s,\ga] =& Ve^{-\s^2}\s + \frac{4N\la^2}{\Om'}\left[\chi\ga^2 + \ga\sqrt{1-\ga^2}\right]^2 e^{-2\s^2}\nonumber\\
& - \frac{\oR}{\s^3} - \frac{Ng}{\sqrt{8\pi}\s^2}\,,
\end{align}
which fixes the width by the condition $F[\s_0(\ga),\ga]=0$.
In the following, we omit the $\ga$-dependence of $\s_0$ and, since the Landau expansion is performed around $\ga=0$, the equilibrium value is understood as $\s_0=\s_0(\ga=0)$.
The zeroth- and second-order expansion coefficients are determined straightforwardly to
\begin{align}
a_0 =& -\frac{\Oa}{2} + \frac{\oR}{2\s_0^2} - \frac{V}{2}e^{-\s_0^2} + \frac{Ng}{\sqrt{8\pi}\s_0}\,,\\
a_2 =& \Oa \left[1-\Biggl(\frac{\la}{\la_\stwo}\Biggr)^2\right]\,.
\end{align}
To evaluate the higher-order Landau coefficients, we first perform the derivatives of the steady-state width $\s_0$ with respect to the order parameter $\ga$.
By means of the theorem of implicit functions, for which we use the auxiliary function $F[\s,\ga]$, we find the implicit derivatives
\begin{align}
\s_0' =& 0\,,\\
\s_0'' =& -8\Oa \left(\frac{4\oR}{\omega_\s^2}\right) \left(\frac{\la}{\la_\stwo}\right)^2 \s_0\,,\\
\s_0''' =& 6\chi\s_0''\,,\\
\s_0^{(4)}=&\left(\frac{3-12\s_0^2}{\s_0}\right)\left(\s_0''\right)^2 - \left(12-6\chi^2\right)\s_0''\,,
\end{align}
where we have defined the frequency of the atomic breathing mode
\begin{equation}
\omega_\s^2 = 4\oR\left[\frac{3\oR}{\s_0^4} + \frac{Ng}{\sqrt{2\pi}\s_0^3} + V(1-2\s_0^2)e^{-\s_0^2}\right]\,.\label{eq:omega_s}
\end{equation}
With these relations, the Landau coefficients $a_n = (\partial_\ga^n E/n!)_{\ga=0}$ up to sixth order are given in the compact form
\begin{align}
a_3 =& -2\Oa\left(\frac{\la}{\la_\stwo}\right)^2\chi\,,\\
a_4 =& \Oa\left(\frac{\la}{\la_\stwo}\right)^2 \left(1+\s_0\s_0''-\chi^2\right)\,,\label{eq:Landau4}\\
a_5 =& \Oa\left(\frac{\la}{\la_\stwo}\right)^2 \left(1+4\s_0\s_0''\right)\chi\,,\\
a_6 =& \frac{\Oa}{6}\left(\frac{\la}{\la_\stwo}\right)^2\left[ (1-4\s_0^2)(\s_0'')^2 - 12\s_0 (1-3\chi^2)\s_0''\right]\,.
\end{align}
As the second derivative of the width $\s_0''$ is always smaller than zero, the sixth-order expansion coefficient consequently fulfills $a_6>0$ for the set of parameters considered in the following other than $\la=0$.

\begin{figure*}
\centering\includegraphics[width=0.75\textwidth]{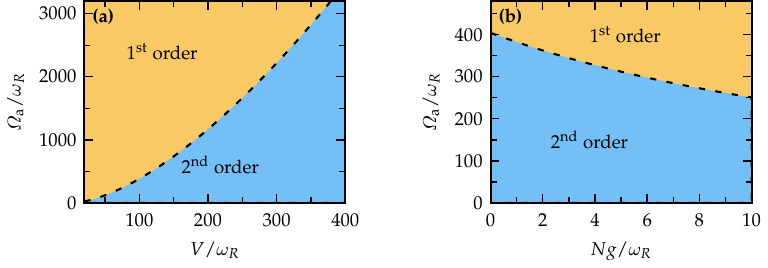}
\caption{The phase diagram of the NQPT is shown as a function of the atomic transition frequency $\Oa$ and (a) the lattice depth $V$ or (b) the interaction strength $Ng$.
While the phase transition is  continuous in the blue regions, it is  discontinuous in the orange regions.
The dashed curves show the critical transition frequency $\Oa=\Omega_\crit$ according to Eq.~\eqref{eq:Ocrit}.
The fixed parameter in (a) is the interaction strength $Ng = \oR$ and in (b) the potential depth $V = 100\,\oR$.}
\label{fig4}
\end{figure*}

In order to describe a first-order nonequilibrium quantum phase transition in the
symmetric coupling regime ($\chi=0$), it is sufficient to consider the Landau coefficients up to sixth order.
In this regime, the odd Landau coefficients vanish, since $a_{2n+1}\sim\chi$ for all $n\in\mathbb{N}$.
In order to quantify the order of the phase transition, we have to look at the sign of the expansion coefficient $a_4$.
To be more specific, the phase transition is continuous when $a_4>0$ and discontinuous for $a_4<0$ at the critical point.
In fact, the fourth-order coefficient exhibits a point at which its sign changes.
From Eq.~\eqref{eq:Landau4}, it follows that this point is given by the relation $1+\s_0\s_0''=0$.
In addition, when $a_4=0$, the phase transition occurs at the critical point $\la=\la_\stwo$.
Hence, we can insert this expression for the atom-membrane coupling rate to find the relation
\begin{equation}
\Omega_\crit = \frac{\omega_\s^2}{32\oR \log(\la_\stwo/\la_\Omega)}\label{eq:Ocrit}\,,
\end{equation}
which defines a critical value for the atomic transition frequency.
Below $\Oa\leq\Omega_\crit$, the phase transition is continuous (second-order) and becomes discontinuous (first-order) for transition frequencies satisfying $\Oa>\Omega_\crit$.
This behavior is depicted in the two panels of Fig.~\ref{fig4}.
In Fig.~\ref{fig4}(a) is shown that by tuning either the potential depth $V$ or the atomic transition frequency $\Oa$, the order of the phase transition can be changed from second- (blue region below dashed curve) to first-order (orange region).
Alternatively, by changing the atomic interaction strength $Ng$, the order may also be tuned, which is shown in Fig.~\ref{fig4}(b).
In that sense, one might also consider to define a critical optical lattice depth $V_\crit$ or critical interaction strength $g_\crit$ (or atom number $N_\crit$) rather than a critical atomic transition frequency. 

The critical coupling rate $\la_\stwo$ in the second-order regime has been derived by simple arguments on the basis of the steady-state equations, which is equivalent to evaluating the lowest-order Landau coefficient $a_2$.
On the other hand, in order to determine the critical coupling rate $\la_\sone$ in the symmetric first-order regime, one has to know at least the expansion coefficients up to sixth order.
For $a_4<0$, the effective nonequilibrium potential $E(\ga)$ exhibits three minima on the real axis, which are located at
\begin{equation}
\ga_1 = 0\,,\qquad \ga_{2,3}^2 = -\frac{a_4}{3a_6} + \sqrt{\left(\frac{a_4}{3a_6}\right)^2 - \frac{a_2}{3a_6}}\,,
\end{equation}
if $a_2>0$, or equivalently $\la<\la_\stwo$.
The local minimum at $\ga_1=0$ has a value of $E(0)=a_0$ and the critical coupling rate $\la_\sone$ is fixed by the condition  $E(\ga_{2,3}) = a_0$. 
After some tedious but straightforward algebra, this leads to the expression $13a_2a_6=4a_4^2$ from which we can derive the critical coupling rate in the first-order regime according to
\begin{align}
 \left[\frac{\Omega_\crit}{\Oa}-\Biggl(\frac{\la_\sone}{\la_\stwo}\Biggr)^2\right]^2 =&\frac{13}{24} \left[1-\Biggl(\frac{\la_\sone}{\la_\stwo}\Biggr)^2\right]\left[12 \frac{\Omega_\crit}{\Oa}\vphantom{\Biggl(\frac{\la_\sone}{\la_\stwo}\Biggr)^2}\right.\nonumber\\&\left. + \Biggl(\log^{-1}\frac{\la_\stwo}{\la_\Omega}-4\Biggr)\Biggl(\frac{\la_\sone}{\la_\stwo}\Biggr)^2 \right]\,.
\end{align}
We recover the result $\la_\sone=\la_\stwo$ in the limit $\Oa=\Omega_\crit$ and find that the first-order phase transition occurs in general at smaller coupling rates than the second-order NQPT, i.e., $\la_\sone<\la_\stwo$ for $\Oa>\Omega_\crit$.

For  asymmetric coupling ($\chi\neq0$), also odd orders in the Landau expansion take a finite value.
This breaks the symmetry in the nonequilibrium potential $E(\ga)$ with respect to $\ga=0$.
In order to estimate the critical coupling rate $\la_\aone$ in the first-order regime, we consider atomic transition frequencies that satisfy $\Oa<(1-\chi^2)\Omega_\crit$ such that $a_4>0$ is always satisfied.
In this case, it is sufficient to take the Landau expansion up to fourth order.
With the same arguments as in the symmetric first-order regime, we derive an expression for the critical coupling rate from the condition $E(\ga_2)=a_0$, where $\ga_2\neq0$ is one of the two points that minimize the effective potential, whereas the other point is the trivial one at $\ga_1=0$.
Hence, we find the relation $4a_2 a_4 - a_3^2 = 0$, which translates to
\begin{equation}
 \left[1-\Biggl(\frac{\la_\aone}{\la_\stwo}\Biggr)^2\right] \left[1-\chi^2 -\frac{\Omega_\crit}{\Oa} \Biggl(\frac{\la_\aone}{\la_\stwo}\Biggr)^2\right] = \Biggl(\frac{\la_\aone}{\la_\stwo}\Biggr)^2 \chi^2\, ,
\end{equation}
which is independent of the sign of $\chi$. 
It is straightforward to show that also $\la_\aone<\la_\stwo$ and $\la_\aone=\la_\stwo$ is recovered in the limit $\chi=0$.

\subsection{Hysteresis in the First-Order Regime}
A characteristic feature of a first-order phase transition is the occurrence of a hysteresis when the atom-membrane coupling $\la$ is adiabatically tuned.
In terms of the nonequilibrium potential, this is included by the existence of two or more local minima.
At a certain coupling rate, these local minima become dynamically unstable and eventually turn into a maximum.
At this point, the system jumps to the neighboring local minimum and remains there until
this minimum becomes unstable. In the following, we consider the two generic cases of a symmetric and an asymmetric coupling to discuss this effect.

\begin{figure*}
\centering\includegraphics[width=0.7\textwidth]{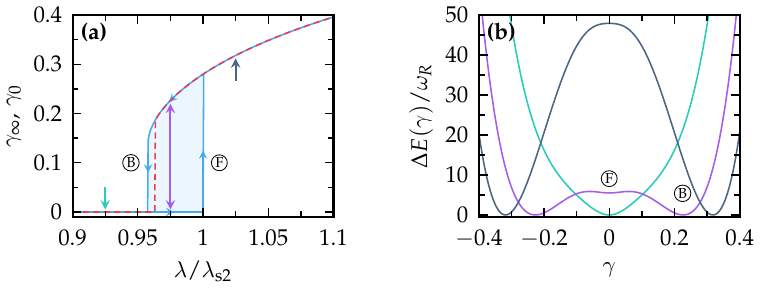}
\centering\includegraphics[width=0.7\textwidth]{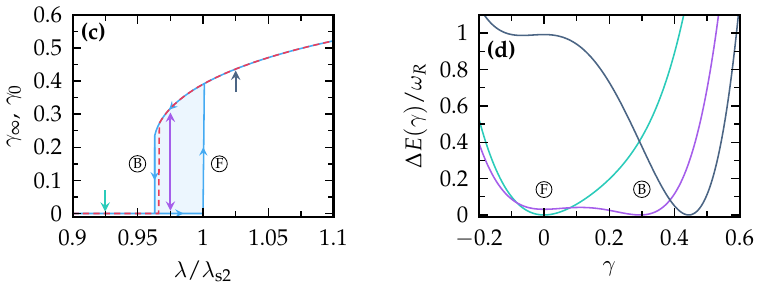}
\caption{Hysteresis curve ($\ga_\infty$ , solid) shown as a function of the coupling parameter \la\ for (a) the symmetric first-order phase transition and (c) the asymmetric first-order phase transition.
The dashed curve shows the globally minimized path $\ga_0$.
(b), (d) Curves of the effective nonequilibrium potential $\Delta E(\ga) = E(\ga )-E(\ga_0)$ are shown for a coupling strength below and above the turning points, $\la\leq \la_\mathrm{B}$ and $\la\geq\la_\mathrm{F}$, and in the coexistence area $\la_\mathrm{B}<\la<\la_\mathrm{F}$.
The colors of the potential curves in (b), (d) mark the associated configurations pointed to by the colored arrows in (a), (c).
The circled letters indicate the minimum of the forward (F) and backward (B) path.
The parameters were chosen as in Fig.~\ref{fig3}(b) and Fig.~\ref{fig3}(c) for the symmetric and asymmetric coupling case, respectively.}
\label{fig5}
\end{figure*}

In order to describe the hysteretic behavior, we take the equations of motion~\eqref{eq:eom-cum1}--\eqref{eq:eom-cum2} with $\ga(t)=\ga_+(t)$ and adiabatically alter the atom-membrane coupling strength.
Thus, we obtain for each value of $\la$ a long-time solution $\ga_\infty = \lim_{t\rightarrow\infty}\ga(t)$ which becomes time independent.
In Fig.~\ref{fig5}(a) and \ref{fig5}(c), we show the hysteresis for the symmetric ($\chi=0$) and asymmetric ($\chi=0.25$) first-order phase transition, respectively.
On the forward path, the coupling strength $\la$ is adiabatically increased and the system is initially prepared in the minimum with the occupation amplitude $\ga=0$.
The system stays there until it becomes unstable at $\la=\la_\mathrm{F}$ and jumps to the adjacent minimum at $\ga\neq0$.
This point coincides with the critical coupling rate $\la_\stwo$ in both of the two cases, the symmetric and asymmetric first-order transition.
Afterwards, the steady-state solution $\ga_0\neq0$ is followed as $\la$ increases.

On the backward path, the system follows the solution with $\ga_0\neq0$ until it becomes dynamically unstable at $\la_\mathrm{B}$ and jumps to the solution at $\ga_0=0$.
For the case of symmetric coupling ($\chi=0$) shown in Fig.~\ref{fig5}(a), this jumping point occurs when $4a_2a_6=a_4^2$.
On the other hand, the jumping point in the regime of asymmetric coupling of Fig.~\ref{fig5}(c) is defined via the relation $32a_2a_4=9a_3^2$.

In the picture of potential energy surfaces, the reason for the hysteretic behavior is the existence of multiple stable local minima at $\ga = 0$ and $\ga\neq 0$ in the coexistence region $\la_\mathrm{B} \leq \la \leq \la_\mathrm{F}$ as indicated in Fig.~\ref{fig5}(b).
Here, the forward and backward minima are indicated by the circled capital letters F and B, respectively. 
The same argumentation applies to the {asymmetric} first-order phase transition.
The structure of the effective potential surface is shown in Fig.~\ref{fig5}(d) for three different values of the atom-membrane coupling $\la$.
The colors of the potential curves in (b) mark the associated configurations pointed to by the colored arrows in (a).

\section{Spectrum of Collective Excitations and Entanglement}
In the preceding section, we have analyzed the equations of motion in the mean-field regime.
We note that the spectrum of collective excitations is included in the equations of motion~\eqref{eq:eom-cum1}--\eqref{eq:eom-cum2}~\cite{Mann2018}.
Yet, we present a different approach that utilizes an adapted Bogoliubov ansatz.
Hence, we start again from the effective equations of motion~\eqref{eq:Langevin1}--\eqref{eq:Langevin2}.

For this purpose, it is convenient to introduce the new field operators
\begin{eqnarray}
\Psi_\NN(z) = \sqrt{1-\ga_0^2} \Psi_-(z) + \ga_0\Psi_+(z)\,,\\
\Psi_\ga(z) = -\ga_0\Psi_-(t,z) + \sqrt{1-\ga_0^2} \Psi_+(z)\,,
\end{eqnarray}
where the first field operator contains the mean-field steady state, and the latter describes excitations out of this steady state via internal transitions.
Let us note that in the steady state, which we are going to determine in the following, the correlation function of the first field operator $\Psi_\NN(z)$ fulfills $\int dz \langle{\Psi_\NN^\dag(z)\Psi_\NN(z)}\rangle=N+N_{\rm qntm}$, where $N_{\rm qntm}$ describes excitations out of the mean-field condensate.
Hence, we chose the label N.
For simplicity, we focus here on the special case of symmetric coupling ($\chi=0$) and non-interacting atoms with $g=0$.
It follows that the equations of motion for these new fields are given by 
\begin{widetext}\begin{align}
 i\partial_t \hat{a} =& (\Om-i\Gm)\hat{a} - \frac{\la}{2}\int dz\,\cos(2z)\Biggl[(1-2\ga_0^2)(\Psi^\dag_\NN\Psi_\ga + \Psi_\ga^\dag \Psi_\NN)+ 2\ga_0\sqrt{1-\ga_0^2}(\Psi_\NN^\dag\Psi_\NN - \Psi_\ga^\dag \Psi_\ga)\Biggr] + i\xi_\mem \,,\label{eq:Bog1}\\
 \begin{split}i\partial_t\Psi_\NN =& \left[-\frac{\Oa}{2}(1-2\ga_0^2) - \oR\partial_z^2 - \frac{V}{2}\cos(2z) - \la \ga_0\sqrt{1-\ga_0^2}(\hat{a}+\hat{a}^\dag)\cos(2z)\right]\Psi_\NN\\
 & + \left[ \Oa \ga_0 \sqrt{1-\ga_0^2} - \frac{\la}{2}(1-2\ga_0^2)(\hat{a}+\hat{a}^\dag)\cos(2z)\right]\Psi_\ga \,,\end{split}\\
 \begin{split}i\partial_t\Psi_\ga  = & \left[\frac{\Oa}{2}(1-2\ga_0^2) - \oR\partial_z^2 - \frac{V}{2}\cos(2z) + \la \ga_0\sqrt{1-\ga_0^2}(\hat{a}+\hat{a}^\dag)\cos(2z)\right]\Psi_\ga\\
 & + \left[ \Oa \ga_0 \sqrt{1-\ga_0^2} - \frac{\la}{2}(1-2\ga_0^2)(\hat{a}+\hat{a}^\dag)\cos(2z)\right]\Psi_\NN \,.\end{split}\label{eq:Bog2}
\end{align}\end{widetext}
Here, taking into account the noise operator $\xi_\mem$ is essential, as correlations exponentially decay with a rate of the order of $\mtc{O}(\Gm)$, meaning that all correlation functions would otherwise vanish in the steady-state regime.
The mean-field steady state $\Psi_-\simeq \sqrt{(1-\ga_0^2)N}\psi_0$, $\Psi_+\simeq\sqrt{N}\ga_0\psi_0$ is completely included in the field operator $\Psi_\NN$.
Hence, we can make the ansatz 
\begin{eqnarray}
\Psi_\NN(t,z) \simeq & \Bigl[\sqrt{N}\psi_0(z) + \hat{d}_\s(t)\psi_2(z)\Bigr]e^{-i\mu t}\,,\\
\Psi_\ga(t,z) \simeq & \hat{d}_\ga(t)\psi_0(z)e^{-i\mu t}\,,
\end{eqnarray}
where $\psi_n(z)$ are the quasi-eigenstates
\begin{equation}
\psi_n(z) = \left(\frac{1}{\pi\s_0^2}\right)^{1/4} \frac{1}{\sqrt{2^n n!}} H_n(z/\s_0) \exp\left(-\frac{z^2}{2\s_0^2}\right)\,,
\end{equation}
with the $n$-th Hermite polynomial $H_n(x)$.
In addition, we assume for the membrane ladder operator a similar superposition of mean-field steady state and quantum fluctuations according to $\hat{a}(t)=\sqrt{N}\a_0 + \hat{d}_\a(t)$.
Here, the operators $\hat{d}_x(t)$ with $x\in\{\a,\ga,\s\}$ follow the usual bosonic algebra, i.e., $\cmt{\hat{d}_x}{\hat{d}^\dag_y}=\delta_{xy}$.

In order to evaluate the spectrum of collective excitations, we determine the equations of motion for the ladder operators $\hat{d}_x$ and linearize with respect to the quantum fluctuations by taking into account only the leading order terms in the atom number $N$.
Hence, we assume that the effective coupling \la\ between the membrane and a single atom  is small, yet the collective coupling $\sqrt{N}\la$ can still be large.
From the set of equations~\eqref{eq:Bog1}--\eqref{eq:Bog2}, we find the equation of motion for the vector of the collective modes $\boldsymbol{\hat{d}}=(\hat{d}_\a,\hat{d}_\ga,\hat{d}_\s,\hat{d}_\a^\dag,\hat{d}_\ga^\dag,\hat{d}_\s^\dag)^t$ according to the Bogoliubov--de Gennes equation
\begin{equation}
i\partial_t \boldsymbol{\hat{d}}(t)
 = 
\boldsymbol{M}(\a_0,\ga_0,\s_0)\boldsymbol{\hat{d}}(t) + i\bxi(t) \,,\label{eq:BdG}
\end{equation}
with the noise vector $\bxi = (\xi_\mem,0,0,-\xi_\mem^\dag,0,0)^t$.
The linear stability matrix $\boldsymbol{M}(\a_0,\ga_0,\s_0)$ is a $6\times6$ linear operator whose eigenvalues $\nu_i$ include the frequencies  and the decay rates of the collective excitations.
The matrix contains the full information of the steady-state solution $(\a_0,\,\ga_0,\,\s_0)$ and can be written as
\begin{equation}
\boldsymbol{M}(\a_0,\ga_0,\s_0)= \left(\begin{array}{cc}
\boldsymbol{H}(\a_0,\ga_0,\s_0) & \boldsymbol{G}(\a_0,\ga_0,\s_0) \\
-\boldsymbol{G}(\a_0,\ga_0,\s_0) & -\boldsymbol{H}^*(\a_0,\ga_0,\s_0)
\end{array}\right)\label{eq:Mstab}
\end{equation}
by defining the $3\times3$-matrices $\boldsymbol{H}$ and $\boldsymbol{G}$.
Here, the matrix $\boldsymbol{H}$ has the diagonal entries
\begin{align}
 H_{11} =& \Om - i\Gm\,,\\
 H_{22} =& \Oa(1-2\ga_0^2) + 2 \sqrt{N}\la\ga_0\sqrt{1-\ga_0^2}(\a_0+\a_0^*)e^{-\s_0^2}\,,\\
 H_{33} =& \frac{2\oR}{\s_0^2} + V_\mathrm{eff} (2-\s_0^2) \s_0^2 e^{-\s_0^2}\,,
\end{align}
with the effective lattice depth $V_\mathrm{eff}=V + 4\Oa(\la/\la_\Omega)^2\ga_0^2(1-\ga_0^2)\exp(-\s_0^2)$, whereas all diagonal entries of $\boldsymbol{G}$ are zero.
The off-diagonal elements of the two matrices couple the individual bare modes with each other and take the values
\begin{eqnarray}
H_{12} = -\frac{\sqrt{N}\la}{2}(1-2\ga_0^2)e^{-\s_0^2}\,,\\
H_{13} = \sqrt{2N}\la \ga_0\sqrt{1-\ga_0^2}\s_0^2 e^{-\s_0^2}\,,\\
H_{23} = {\sqrt{2N}\la}(1-2\ga_0^2)\mathrm{Re}(\a_0)\s_0^2 e^{-\s_0^2}\,,
\end{eqnarray}
with $G_{1,i}=H_{1,i}$ for $i\neq1$, and $\boldsymbol{H}^t=\boldsymbol{H}$, $\boldsymbol{G}^t=\boldsymbol{G}$.
All other entries vanish.

\subsection{Collective Excitations and Atom-Membrane Entanglement}
\begin{figure*}
\centering\includegraphics[width=\textwidth]{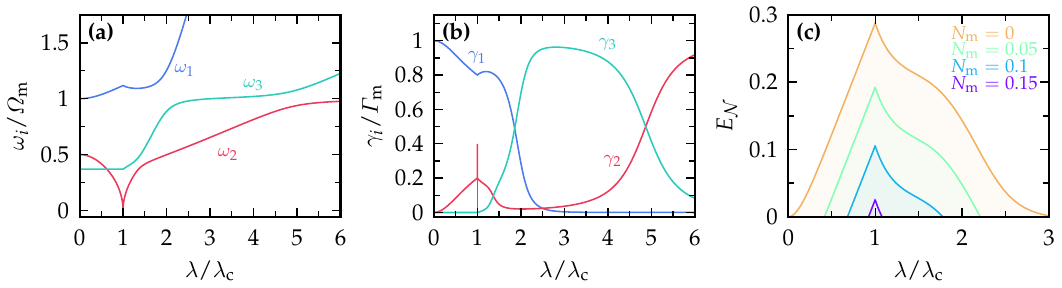}
\caption{(a) The three collective excitation frequencies $\omega_i = \mathrm{Re}(\nu_i)$ are shown as a function of the atom-membrane coupling rate $\la/\la_\crit$ in the weak damping limit. 
The different colors correspond to different eigenmodes.
(b) In addition, the decay rates of the collective modes $\ga_i = -\mathrm{Im}(\nu_i)$ are shown.
We note that due to the weak membrane damping, avoided energy crossings appear, which is also reflected in the rich structure of the collective decay rates.
(c) The logarithmic negativity $E_\mathcal{N}$ is shown as a function of the atom-membrane coupling for different bath occupation numbers $N_\mem$.
Each curve characterizes the entanglement between the vibrational mode of the membrane and the atomic transition mode which is maximized at the critical point $\la=\la_\crit$. 
The logarithmic negativity is strongly reduced as the temperature of the environment is raised.
The parameters used are $V = 100\,\oR$, $\Om=100\,\oR$, $\Gm = \oR$, $\Oa=50\,\oR$, and $g = 0$, $\chi = 0$.
}
\label{fig7}
\end{figure*}

In Fig.~\ref{fig7}, we show the spectrum of the collective excitations in the regime of the continuous phase transition for a weakly damped membrane $\Gm=0.01\,\Om$.
The excitation frequencies of the collective modes $\omega_i=\mathrm{Re}(\nu_i)$ are shown in panel (a).
Below the critical coupling rate $\la<\la_\crit$, the breathing mode $\omega_3$ (green) is constant, while the low energy excitation frequency (red) exhibits a mode softening and monotonically decreases to zero according to $\omega_{2}\simeq\Oa\sqrt{1-(\la/\la_\crit)^2}$. 
Simultaneously, the high-energy frequency $\omega_1$ (blue) monotonically increases.
Above the threshold $\la>\la_\crit$, the low-energy mode frequency increases again.
This mode eventually saturates to \Om , which means that the membrane mode decouples from the atomic modes.
The high-energy frequency $\omega_1$ decreases in a short interval after the critical point.
At the end of this interval, an avoided energy crossing between $\omega_1$ and $\omega_3$ occurs and, afterwards, the two mode frequencies follow a monotonic increase.
The presence of the avoided crossing is a direct consequence of the comparably weak membrane damping rate $\Gm=0.01\,\Om$.
For the sake of completeness, we show the decay rates $\ga_i = -\mathrm{Im}(\nu_i)$ in Fig~\ref{fig7}(b).
In the vicinity of the avoided crossings in (a), the initially small decay rates are significantly increased, which indicates a strong mixing of the membrane mode and the atomic modes.
Moreover, the decay rate $\gamma_2$ exhibits a bifurcation, which appears as a vertical line at the critical coupling rate, where the corresponding excitation frequency $\omega_2$ vanishes.
In fact, a zoom shows that the decay rate bifurcates in a finite interval around the critical point, e.g., see Ref.~\cite{Nagy2011}.

In order to determine the atom-membrane entanglement in the steady-state regime, we solve the Bogoliubov--de Gennes equation~\eqref{eq:BdG} in the long-time limit $t\rightarrow\infty$.
A possible measure to quantify the quantum entanglement between two different bare modes is the logarithmic negativity $E_\mtc{N}$~\cite{Vidal2002,Nagy2011,Adesso2004,Arani2016}, 
which is evaluated from the reduced covariance matrix ${C}_{kl}=\mean{\hat{x}_k \hat{x}_l + \hat{x}_l \hat{x}_k}/2$. 
Here, we adopted the shorthand notation $\boldsymbol{\hat{x}}=(\hat{q}_\a,\hat{q}_\ga,\hat{q}_\s,\hat{p}_\a,\hat{p}_\ga,\hat{p}_\s)^t$ with the quadratures $\hat{q}_\mu = (\hat{d}_\mu +\hat{d}_\mu^\dag)/\sqrt{2}$, $\hat{p}_\mu=(\hat{d}_\mu - \hat{d}_\mu^\dag)/\sqrt{2}i$.
The columns and rows of the irrelevant modes are neglected.
The mean value has to be taken for the long-time solution such that $\mean{\hat{a}\hat{b}}\equiv\lim_{t\rightarrow\infty}\mean{\hat{a}(t)\hat{b}(t)}$.
In general, the covariance matrix $\boldsymbol{C}$ takes the form
\begin{equation}
\boldsymbol{C} = \left( \begin{array}{cc}\boldsymbol{U} & \boldsymbol{V} \\ \boldsymbol{V}^t & \boldsymbol{W}\end{array}\right)\,.
\end{equation}
The logarithmic negativity is then related to the smallest symplectic eigenvalue $\tilde{\nu}_-$ of $\boldsymbol{C}$ and is expressed as $E_\mathcal{N} = \mathrm{max}\left\{0,-\log(2\tilde{\nu}_-)\right\}$ with
\begin{equation}
\tilde{\nu}_- = 2^{-1/2}\sqrt{\Sigma(\boldsymbol{C}) - \sqrt{\Sigma(\boldsymbol{C}) - 4 \det\boldsymbol{C}}}\,,
\end{equation}
and $\Sigma(\boldsymbol{C})=\det\boldsymbol{U}+\det\boldsymbol{W}-2\det\boldsymbol{V}$.
We note that the symplectic eigenvalues of the matrix $\boldsymbol{C}$ corresponds to the eigenvalues of the matrix $\boldsymbol{A}=i\boldsymbol{J}\boldsymbol{C}$, where $\boldsymbol{J}$ is the skew-symmetric matrix \begin{equation}\boldsymbol{J}=\left(\begin{array}{cc}
0 & \boldsymbol{1} \\
-\boldsymbol{1} & 0
\end{array}\right)\,,\end{equation}
and $\boldsymbol{1}$ is the $2\times2$ identity matrix.

The resulting entanglement between the membrane mode $\hat{d}_\a$ and the atomic transition mode $\hat{d}_\ga$ is shown in Fig.~\ref{fig7}(c).
The logarithmic negativity $E_\mathcal{N}$ exhibits a non-differentiable, global maximum at the critical coupling rate $\la=\la_\crit$.
Above the critical point, the entanglement is progressively reduced as the atom-membrane coupling increases.
This behavior intermediately flattens out in the region of the avoided crossings, which are indicated in Fig.~\ref{fig7}(a) around $\la\simeq1.5\,\la_\crit$.
The differently colored curves correspond to cases with different environmental occupation numbers $N_\mem$, which enter in the autocorrelation function of the noise in Eq.~\eqref{eq:autocorr}.

\subsection{Excitation Spectrum Along the Hysteresis}

In the regime of a first-order phase transition $\Oa>\Omega_\crit$, the experimentally observed spectrum of excitations does, in general, not coincide with the predicted excitation spectrum of equation \eqref{eq:Mstab}.
This is because the globally minimized solution $(\a_0,\,\ga_0\,,\s_0)$ differs from the long-time solution $(\a_\infty,\,\ga_\infty,\,\s_\infty)$ in the vicinity of the critical point $\la_\crit$ due to the presence of a hysteresis.
In the following, we, therefore, distinguish between the \textit{minimal} spectrum, which is evaluated with the globally minimized state $(\a_0,\ga_0,\s_0)$, and the \textit{forward} (\textit{backward}) spectrum, which is determined with the long-time solution $(\a_\infty,\,\ga_\infty,\,\s_\infty)$ of the equations of motion \eqref{eq:eom-cum1}--\eqref{eq:eom-cum2} for a coupling strength $\la$ that is adiabatically raised (reduced).
In order to determine the spectrum of collective excitations along the hysteresis of a symmetric first-order phase transition, we evaluate the linear stability matrix $\boldsymbol{M}(\a_\infty,\ga_\infty,\s_\infty)$.

\begin{figure*}
\centering\includegraphics[width=0.9\textwidth]{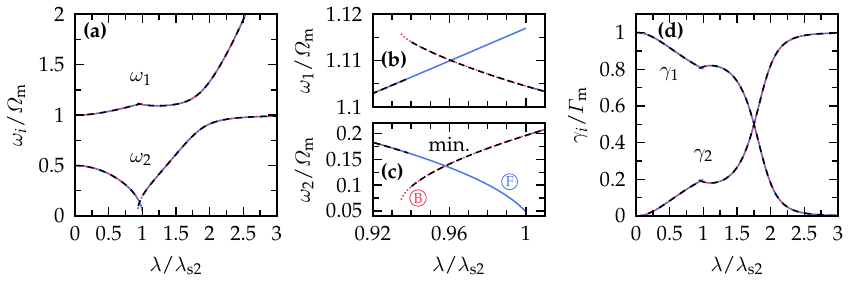}
\caption{(a) The high-energy excitation frequency $\omega_1=\mathrm{Re}(\nu_1)$ and the mode softening-type excitation frequency $\omega_2=\mathrm{Re}(\nu_2)$ are shown along the hysteresis in the regime of a symmetric first-order phase transition, i.e., $\chi=0$.
Slight deviations between the globally minimized solution and the forward/backward propagation are found only in the vicinity of the critical point $\la=\la_\crit$ and the coupling strenght $\la=\la_\stwo$.
A zoom of (b) $\omega_1$ and (c) $\omega_2$ around $\la_\crit$ visualizes these deviations.
Curves with different colors correspond to different solutions indicated in (c), where F (B) corresponds to the forward (backward) solution of the hysteresis with values $(\a_\infty,\ga_\infty,\s_\infty)$ and min. is the globally minimized solution with values $(\a_0,\ga_0,\s_0)$.
(d) The corresponding decay rates $\ga_1$ and $\ga_2$ of the collective eigenmodes are shown along the hysteresis.
The other parameters chosen are $V = 100\,\oR$, $\Om=10^5\,\oR$, $\Gm = 0.01\,\Om$, $\Oa=0.5\,\Om$, and $g = 0$.
}
\label{fig8}
\end{figure*}

The high-energy excitation frequency $\omega_1$ and the mode softening-type excitation frequency $\omega_2$ are shown in Fig.~\ref{fig8}(a) as a function of the atom-membrane coupling strength $\la$.
Here, the solid (blue) curve illustrates the \textit{forward} solution, the dotted (red) curve the \textit{backward} solution and the dashed (black) curve shows the excitation frequency in the \textit{minimal} state.
Deviations between the curves are only present when the steady-state solution $(\a_0,\ga_0,\s_0)$ and the long-time solution $(\a_\infty,\ga_\infty,\s_\infty)$ differ from each other.
For a better visualization, we show a zoom around the critical point $\la=\la_\crit$ in Fig.~\ref{fig8}(b) and (c).
Following the \textit{forward} path of the softening mode frequency $\omega_2$ in (c), it coincides with the dashed, \textit{minimal} solution evaluated up to the critical coupling $\la_\mathrm{s1}$, then further decreases until it reaches a minimal value at $\la=\la_\mathrm{s2}$ and jumps back to the \textit{minimal} solution.
A similar behavior is found for the \textit{backward} solution in the opposite direction and the high-energy excitation frequency $\omega_1$ in (b).
For completeness, we show the corresponding decay rates of the collective excitation modes $\ga_1$ and $\ga_2$ in Fig.~\ref{fig8}(d).
In contrast to the second-order phase transition, no bifurcation of the decay rate of the low-energy excitation mode $\ga_2$ is observed.
We note that due to the large frequency mismatch between the breathing mode and the remaining  modes, the breathing mode decouples and is very well approximated by $\nu_3\simeq \omega_\s=4\sqrt{\oR V_\mathrm{eff}(1-2\s_0^2)e^{-\s_0^2}}$ and therefore omitted from the figure.

\section{Experimental Realization}
Current experimental set-ups use the motional coupling scheme~\cite{Vogell2013,Joeckel2015}.
In order to observe a first-order phase transition in this set-up, one has to compare two different energy scales with each other, one which corresponds to the breathing mode frequency of the condensate (here $\omega_\s$) and the other which is of the order of the energy difference between the normal and the symmetry-broken phase (here $\Delta E \sim \Oa$).
In the motional coupling scheme, the symmetry-broken phase is a configuration where the atomic center-of-mass position is displaced with respect to the lattice minima defined by the optical potential with depth $V$.
Hence, the relevant energy difference scales with frequency of the displacement mode that is given by $\omega_{\zeta}=\sqrt{4\oR V}e^{-\s_0^2}$~\cite{Mann2018}.
Accordingly, both $\omega_\s$ and $\omega_\zeta$ do not scale independently and a first-order phase transition cannot be observed.
The internal state coupling scheme overcomes this limitation.
The direct observation is possible by either measuring the membrane eigenfrequency or the condensate width $\sigma$~\cite{Mann2018}.
In the case of a first-order NQPT, these quantities exhibit a jump at the critical coupling rate, rather than a continuous behavior as in the case of a second-order NQPT, see Figs.~\ref{fig7} and~\ref{fig8}.
Moreover, a direct measurement of the condensate occupation amplitude $\ga_0$ ($\ga_\infty$) can detect the NQPT in a straightforward way.
Furthermore, from a quantum information perspective, the internal state coupling scheme is superior to the motional coupling scheme, because the information can be stored in discrete atomic states rather than continuous, motional states.

For instance, realistic experimental parameters $\Om=70\,\oR$ \cite{Zhong2017,Christoph2018}, $\Oa=20\,\oR$, $\chi\simeq1$ lead to a critical 
coupling rate of $\sqrt{N}\la_\crit\simeq29\,\oR$, which can be reached by increasing the cavity finesse or the laser power of current experiments by a factor of 10.
Within this consideration, the internal states correspond to the $^{87}$Rb hyperfine
states $\ket{-}=\ket{5^2S_{1/2},F=2,m_f=2}$, $\ket{+}=\ket{5^2S_{1/2},F=2,m_f=0}$ and the excited state $\ket{\mathrm{e}}=\ket{5^2P_{1/2},F=2,m_f=1}$~\cite{Steck2009}.

Furthermore, the asymmetry parameter $\chi$ can be tuned by applying an additional perpendicular laser field that drives the transition from \ket{-} to \ket{+}, giving rise to a term $\delta(\ga_+^* \ga_- + \ga_-^*\ga_+)$ in the potential energy \eqref{eqn:Epot}.
Compensating  an additional force on the membrane that scales with $\sqrt{N}\la$ and tuning the parameter $\delta$ allows for an indirect variation of $\chi$.

\section{Conclusion}
We have shown that the hybrid atom-optomechanical system not only undergoes a nonequilibrium quantum phase transition between phases of different collective behavior, but also that the order of the phase transition can be tuned in a straightforward manner.
The steady state of an atomic condensate in an optical lattice, whose internal states are coupled to a single mechanical vibrational mode of a distant membrane, has
been analyzed, based on a Gross--Pitaevskii-like mean-field approach with a time-dependent Gaussian variational ansatz.
Mediated by the light field of a common laser, the atom-membrane coupling is tuned by changing the laser intensity.
Below a critical coupling $\la_\crit$, all the atoms occupy the energetically lower state \ket- and at the critical point a nonequilibrium quantum phase transition occurs.
This phase is characterized by a sizeable steady-state occupation of the energetically higher state \ket+ and a constantly displaced membrane.
The order of this nonequilibrium quantum phase transition is determined by the state-dependent atom-membrane coupling and the transition frequency $\Oa$.
For an asymmetric coupling, $\chi\neq0$, an asymmetric first-order phase transition occurs with a preferred occupation amplitude of the internal states.
Instead, for a symmetric coupling, $\chi = 0$, the phase transition is discontinuous for transition frequencies above a critical value $\Omega_\crit$.
Moreover, the first-order phase transition is accompanied by hysteresis.
On the other hand, when the transition frequency $\Oa$ is smaller than the critical value, the $U(1)$-symmetry of the internal states is spontaneously broken and a second-order phase transition occurs.
This phase transition is characterized by an enhanced atom-membrane entanglement, a mode softening of the frequency and a bifurcation of the decay of the low-energy excitation mode.
The transition between a first- and second-order is observable by tuning readily accessible parameters in the internal state coupling scheme.

\section{acknowledgement}
This work was supported by the
Deutsche Forschungsgemeinschaft via an individual project (N.M., M.T., project number 274978739) and the Collaborative Research Center SFB/TR185 (A.P., project number 277625399).

\end{document}